\documentclass[prl,aps,amssymb,amsmath,tightenlines,twocolumn]{revtex4}

\usepackage{graphicx,epsfig,xcolor}
\usepackage[T1]{fontenc}
\usepackage[latin1]{inputenc}
\usepackage[english]{babel}
\usepackage{amsmath}
\usepackage{mathtools}
\usepackage{amsfonts}
\usepackage{amssymb}
\usepackage{dsfont}

\newcommand{\cL}{\ensuremath{\mathcal{L}}}
\newcommand{\cD}{\ensuremath{\mathcal{D}}}
\newcommand{\cPT}{\ensuremath{\mathcal{PT}}}
\newcommand{\half}{\mbox{$\textstyle{\frac{1}{2}}$}}

\newcommand{\threefourth}{\mbox{$\textstyle{\frac{3}{4}}$}}
\newcommand{\third}{\mbox{$\textstyle{\frac{1}{3}}$}}
\newcommand{\tthird}{\mbox{$\textstyle{\frac{2}{3}}$}}
\newcommand{\fourth}{\mbox{$\textstyle{\frac{1}{4}}$}}
\newcommand{\fifth}{\mbox{$\textstyle{\frac{1}{5}}$}}
\newcommand{\sixth}{\mbox{$\textstyle{\frac{1}{6}}$}}

\begin{document}

\title{Underdetermined Dyson-Schwinger equations}
\author{Carl M.~Bender$^a$}\email{cmb@wustl.edu}
\author{Christos
Karapoulitidis$^b$}\email{christos.karapoulitidis@stud.uni-heidelberg.de}
\author{S.~P.~Klevansky$^b$}\email{spk@physik.uni-heidelberg.de}

\affiliation{
$^a$Department of Physics, Washington University, St.~Louis, Missouri 63130,
USA\\
$^b$Institut f\"ur Theoretische Physik, Universit\"at Heidelberg, 69120
Heidelberg, Germany\\ }

\begin{abstract}
This paper examines the effectiveness of the Dyson-Schwinger (DS) equations as
a calculational tool in quantum field theory. The DS equations are an infinite
sequence of coupled equations that are satisfied exactly by the connected
Green's functions $G_n$ of the field theory. These equations link lower to
higher Green's functions and, if they are truncated, the resulting finite system
of equations is underdetermined. The simplest way to solve the underdetermined
system is to set all higher Green's function(s) to zero and then to solve the
resulting determined system for the first few Green's functions. The $G_1$ or
$G_2$ so obtained can be compared with exact results in solvable models to see
if the accuracy improves for high-order truncations. Five $D=0$ models are
studied: Hermitian $\phi^4$ and $\phi^6$ and non-Hermitian $i\phi^3$, $-\phi^4$,
and $i\phi^5$ theories. The truncated DS equations give a sequence of
approximants that converge slowly to a limiting value but this limiting value
always {\it differs} from the exact value by a few percent. More sophisticated
truncation schemes based on mean-field-like approximations do not fix this
formidable calculational problem.
\end{abstract}
\maketitle

The objective in quantum field theory is to calculate {\it connected Green's
functions} $G_n(x_1,... x_n)$, which contain the physical content of the theory.
In principle, the program is to solve the field equations for the field
$\phi(x)$ and then to calculate vacuum expectation values (VEVs) of
time-ordered products of $\phi$:
$\gamma_n(x_1,...x_n)\equiv\langle 0|{\rm T}\{\phi(x_1) ... \phi(x_n)\}
|0\rangle.$
The nonconnected Green's functions $\gamma_n$ are then combined into {\it
cumulants} to get $G_n$ \cite{r1}.

The Dyson-Schwinger (DS) equations purport to be a way to calculate both the
perturbative and nonperturbative behavior of $G_n$ by using c-number functional
analysis without resorting to operator theory \cite{coleman, r2,r3,r4}. The 
procedure is to truncate the infinite system of coupled DS equations to a 
finite set of coupled equations that would give good approximations to the 
first few $G_n$. The problem is that, while the DS equations are satisfied exactly 
by $G_n$, the DS equations are an {\it underdetermined} system; each new 
equation introduces additional Green's functions, so a truncation of the system 
contains more Green's functions than equations \cite{oldpaper}. A plausible 
strategy is to close the truncated system by setting the highest Green's function(s) 
to zero and then to solve the resulting determined system.

Here we study the simplest case: quantum field theory in zero-dimensional
spacetime. Successive elimination gives {\it polynomial} equations for $G_1$ or
$G_2$. We examine the convergence and accuracy of this procedure as the
system of coupled equations increases in size for five $D=0$ theories, Hermitian
quartic and sextic theories and non-Hermitian $\cPT$-symmetric cubic, quartic,
and quintic theories \cite{r5}. The truncated DS equations provide fair
numerical values for the connected Green's functions, but these approximations
do not converge to the exact results when they are examined in high order.

The DS equations follow
directly on differentiating the functional integral for $Z[J]$ (or $\log(Z[J])$)
with respect to $J$, giving $\gamma_n$ (or $G_n$), 
$$Z[J]=\textstyle{\int\cD\phi\,\exp\int dx}\{-\cL[\phi(x)]+J(x)\phi(x)\},$$
where $\cL$ is the Lagrangian, $J$ is a  c-number source, and $Z[0]$ is the 
Euclidean partition function \cite{bms, r7}.

\vspace{0.1cm}
\noindent{\bf Hermitian quartic $D=0$ theory:} The functional
integral $Z[J]$ becomes an ordinary integral $Z[J]=\int_{-\infty}^\infty d\phi\,
e^{-\cL(\phi)}$, where $\cL(\phi)=\fourth\phi^4-J\phi$. The exact connected
two-point Green's function is:
\begin{eqnarray}
G_2&=&\textstyle{\int_{-\infty}^\infty}d\phi\,\phi^2 e^{-\phi^4/4} \big/
\textstyle{\int_{-\infty}^\infty}d\phi\,e^{-\phi^4/4}\nonumber\\
&=& 2\Gamma\big(\threefourth\big)\big/\Gamma\big(\fourth\big)= 0.675\,978\,240
...\,.
\label{e5}
\end{eqnarray}

We impose parity invariance when $J=0$, so all odd Green's functions vanish.
The first nontrivial DS equation reads $G_4=-3G_2^2+1$. Truncating this equation 
by setting $G_4=0$, we obtain the approximate result $G_2=1/\sqrt{3}=0.577\,350...\,$.
In comparison with (\ref{e5}), this result is 14.6\% low, which is unimpressive.

The next three DS equations are
\begin{eqnarray}
G_6 \!&=& \!-12G_2G_4-6G_2^3,\nonumber\\
G_8 \!&=& \!-18G_2G_6-30G_4^2-60G_2^2G_4,\\
G_{10}\!&=&\!-24G_2G_8-168G_4G_6-126G_2^2G_6-420G_2G_4^2.\nonumber
\label{e6}
\end{eqnarray}
This system is underdetermined; the number of unknowns is always one more than
the number of equations. To solve this system we eliminate $G_4$ by substituting
the first equation into the second, we eliminate $G_6$ by substituting the first
two equations into the third, and so on. We obtain $G_{2n}$ as an $n$th degree
polynomial $P_n(G_2)$ (dividing by the coefficient of the highest power of
$G_2$):
\begin{eqnarray}
\!\!\!\!\!\!\!\!&&\!\!\!\!\!\!P_2(x)=x^2-\third,~P_3(x)=x^3-\textstyle{
\frac{2}{5}}x,
\nonumber\\
\!\!\!\!\!\!\!\!&&\!\!\!\!\!\!P_4(x)=x^4-\textstyle{\frac{8}{15}}x^2+\textstyle{\frac{1}{21}},~
P_5(x)=x^5-\tthird x^3+\textstyle{\frac{193}{1890}}x.
\label{e7}
\end{eqnarray}

Closing the truncated DS equations means finding the zeros of these polynomials.
The positive roots are plotted in Fig.~\ref{f1}. These roots are real and
nondegenerate, and range upwards towards the exact $G_2$ in (\ref{e5}). We
cannot know {\it a priori} which root best approximates $G_2$ but the roots
become denser at the upper end, so we would guess that the largest root gives
the best approximation.

\begin{figure}[t]
\centering
\includegraphics[scale = 0.29]{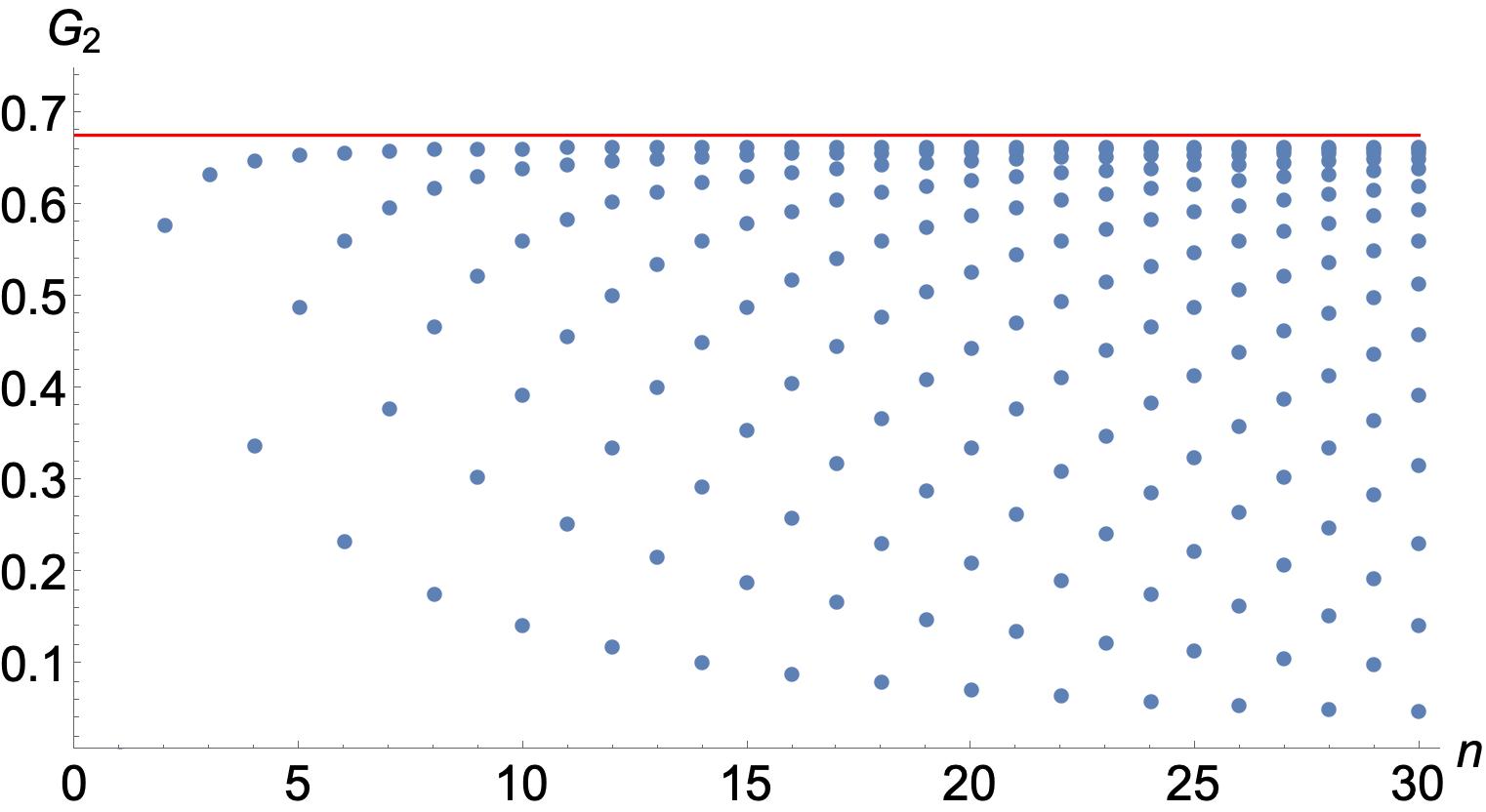}
\caption{Positive zeros of $P_n(x)$ in (\ref{e7}) plotted as a function of $n$
up to $n=30$. The zeros are nondegenerate and range from 0 up to just below
the exact value of $G_2=0.675978...$ (\ref{e5}) (heavy horizontal line).
}
\label{f1}
\end{figure}

\vspace{0.1cm}
\noindent{\bf Inaccuracy of DS approximants:} The accuracy of the largest root
in Fig.~\ref{f1} improves slowly and monotonically with the order of the
truncation. However, while the sequence of largest roots in Fig.~\ref{f1}
converges as $n\to\infty$, the limiting value is $0.663488...$, which is $1.85\%$
{\it below} the exact value of $G_2$ in (\ref{e5}). This discrepancy arises
because truncating the DS equations means replacing $G_{2n}$ by $0$, but
$G_{2n}$ is {\it not small}. The DS equations are exact, so we can compute
$G_{2n}$ by substituting $G_2$ in (\ref{e5}) into (\ref{e6}). We find that
the Green's functions grow rapidly with n: $G_{20}= -4.2788\times 10^9$, $G_{22}
=3.0137\times 10^{11}$. Richardson extrapolation \cite{r6} yields the
asymptotic behavior of $G_{2n}$:
\begin{equation}
G_{2n} \sim 2r^{2n}(-1)^{n+1}(2n-1)!~~(n\to\infty),
\label{e8}
\end{equation}
where $r=0.409\,505\,7...\,.$

Because the DS equations are algebraic when $D=0$, we can derive this asymptotic
behavior analytically: We substitute $G_{2n}=(-1)^{n+1}(2n-1)!\,g_{2n}$,
multiply the $2n$th DS equation by $x^{2n}$, sum from $n=1$ to $\infty$, and
define the generating function $u(x)\equiv x g_2+x^3 g_4+x^5 g_6 +...\,$. The
differential equation satisfied by $u(x)$ is nonlinear:
\begin{equation}
u''(x)=3u'(x)u(x)-u^3(x)-x,
\label{e9}
\end{equation}
where $u(0)=0$ and $u'(0)=G_2$. We {\it linearize} (\ref{e9}) by substituting
$u(x)=-y'(x)/y(x)$ and get $y'''(x)=xy(x)$, where $y(0)=1$, $y'(0)=0$,
$y''(0)=-G_2$. The exact solution satisfying these initial conditions is
\begin{equation}
y(x)=\textstyle{\frac{2\sqrt{2}}{\Gamma(1/4)}\int_0^\infty}dt\,\cos(xt)\,
e^{-t^4/4}.
\label{e11}
\end{equation}
If $y(x)=0$, the generating function $u(x)$ becomes infinite, so the smallest
value of $|x|$ at which $y(x)=0$ is the radius of convergence of the series for
$u(x)$. A simple plot shows that $y(x)$ vanishes at $x_0=\pm 2.441\,968\,2...
$\cite{r7}. Therefore, $r=1/x_0=0.409\,506...\,$, which confirms (\ref{e8}).

The asymptotic behavior in (\ref{e8}) indicates that $G_{2n}$ grows {\it much
faster} than the $\gamma_{2n}$ as $n\to\infty$:
$$\gamma_{2n}=\textstyle{\frac {\int_{-\infty}^\infty dx\,x^{2n}e^{-x^4/4}}
{\int_{-\infty}^\infty dx\,e^{-x^4/4}}\sim 2^n\frac{\Gamma(n/2+1/4)}
{\Gamma(1/4)}}.$$
This is astonishing because we get the connected Green's functions by {\it
subtracting} the disconnected parts from $\gamma_{2n}$. 

Surprisingly, neglecting the huge quantity $G_{2n}$ on the left side of the DS
equations (\ref{e6}) still leads to a reasonably accurate result for $G_2$, as
Fig.~\ref{f1} shows. This is because while the term on the left side is big, the
terms on the right are comparably big \cite{r7}. We also find that Pad\'e 
approximants or mean-field-like schemes do {\it not} improve the convergence. 
But there {\it is} a way to get accurate results: Approximating the left side of the
DS equations with the asymptotic formula in (\ref{e8}) gives $G_2$ to high
precision (see Fig.~\ref{f3}. This approach 
works well for $D=0$ but is difficult to implement if $D>0$ as the DS equations are coupled nonlinear 
integral equations instead of algebraic equations.  
\begin{figure}[t]
\centering
\includegraphics[scale = 0.24]{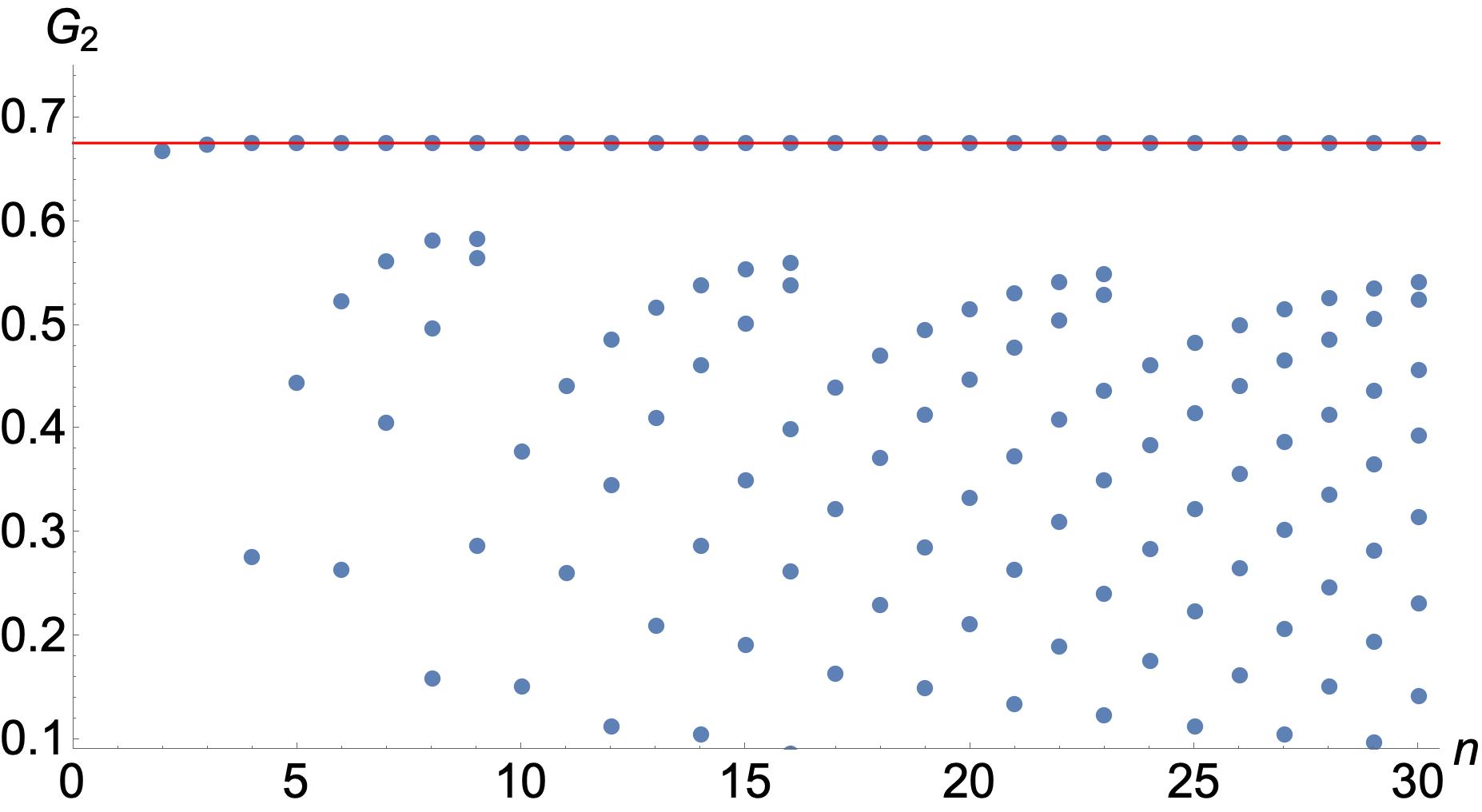}
\caption{Dramatic improvement of the results in Fig.~\ref{f1} obtained by
replacing the left side of the DS equations (\ref{e6}) by the asymptotic
approximation (\ref{e8}) instead of zero. 
}
\label{f3}
\end{figure}

\vspace{0.1cm}
\noindent{\bf Non-Hermitian cubic $D=0$ theory:} The massless Lagrangian
$\cL=\third i\phi^3$ defines a non-Hermitian $\cPT$-symmetric theory whose
one-point Green's function is
\begin{equation}
G_1=\textstyle{\int}dx\,x e^{-ix^3/3}\big/\textstyle{\int}dx\,e^{-ix^3/3},
\label{e12}
\end{equation}
where the path of integration terminates in a $\cPT$-symmetric pair of Stokes
sectors \cite{r5}, so the exact value of $G_1$ is $G_1=-i3^{1/3}\Gamma
\big(\tthird\big)\big/\Gamma\big(\third\big)=-0.729\,011\,13...\,i.$

The first four DS equations are 
\begin{eqnarray}
\!\!\!&&\!\!\!\!\!\!\!\!G_2=-G_1^2,~G_3=-2G_1G_2-i,\nonumber\\
\!\!\!&&\!\!\!\!\!\!\!\!G_4=-2G_2^2-2G_1G_3,~G_5=-6G_2G_3-2G_1G_4.
\label{e13}
\end{eqnarray}
To obtain the leading approximation to $G_1$ we substitute the first equation
into the second and truncate by setting $G_3=0$. The resulting equation is
$G_1^3=\half i$ and the solution that is consistent with $\cPT$ symmetry is
$G_1=-2^{-1/3}i=-0.793\,700\,53...\,i$. This result differs by $8.9\%$ from the
exact value of $G_1$.

At higher order we again truncate the system and find the roots of the associated
polynomial in $G_1$. At first, the roots consistent with $\cPT$ symmetry obtained 
by this procedure approach the exact $G_1$ but unlike the roots for the Hermitian 
quartic theory, where the approach is monotone (Fig.~\ref{f1}), the approach is 
oscillatory: For the $n=4,5,6,7$ truncations the closest roots to the exact $G_1$ are $-0.693\,
361...\,i$, $-0.746\,900...\,i$, $-0.712\,564...\,i$, and $-0.739\,871...\,i$.
However, for $n=8$ this pattern breaks; the closest root is $-0.712\,368...\,i$,
which is a worse approximation than the $n=6$ root.

\begin{figure}[h]
\centering
\includegraphics[scale = 0.13]{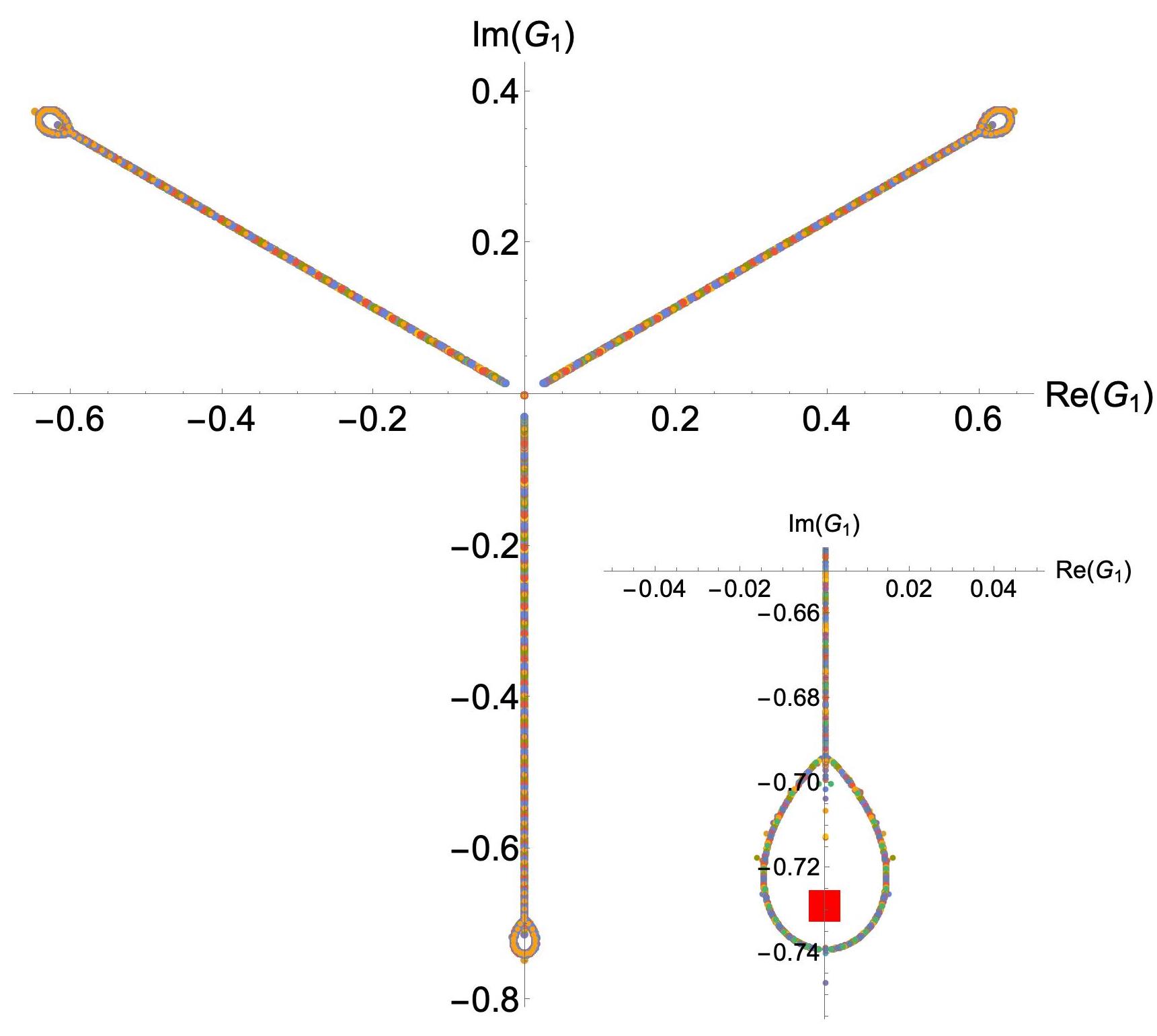}
\caption{All solutions of the truncated DS equations (\ref{e13}) for the
non-Hermitian cubic theory. Inset: The square indicates the exact
$G_1=-0.729\,011\,13...\,i$. 
}
\label{f4}
\end{figure}

This departure from oscillatory convergence is the first indication of a
qualitative change in the approximants. For $n=10$ the roots closest to $G_1$
are a pair on either side of the negative-imaginary axis at $-0.717\,367...\,i
\pm 0.016\,050...\,$. We solve the DS equations up to the 150th truncation and
plot in Fig.~\ref{f4} all roots from $n=2$ to 150 as dots in the complex plane.
These roots become dense on a three-bladed propeller shape, with a small loop at
the tip of each blade. The inset shows that dots on the loop surround but do not
approach the exact $G_1$.

The roots in Fig.~\ref{f4} have threefold symmetry because the truncated DS
equations give polynomials having only powers of $x^3$ (apart from a root at 0).
The DS equations depend only {\it locally} on the integrand of the functional
integral; they are totally insensitive to the boundary conditions on the
functional integrals. There are three pairs of Stokes sectors of angular opening
$60^\circ$ inside of which the integration path in (\ref{e12}) can terminate.
These sectors are centered about $\theta_1=\frac{\pi}{2}$, $\theta_2=-\frac{\pi}
{6}$, or $\theta_3=-\frac{5\pi}{6}$. If the integration path terminates in the
\cPT-symmetric (2,3) sectors, $G_1$ is negative imaginary, but if it terminates
in the (1,2) or (1,3) sectors, $G_1$ is complex.

\vspace{0.1cm}
\noindent{\bf Asymptotic behavior of $G_n$ for large $n$:} 
Richardson extrapolation gives the large-$n$ behavior of the exact Green's
functions for the cubic theory ($G_{14}=42\,692.806\,116$, $G_{15}=-255\,589.034\,701 \,i$):
\begin{equation}
G_n\sim-(n-1)!\,r^n(-i)^n\quad(n\to\infty),
\label{e14}
\end{equation}
where $r=0.427\,696\,347\,707...\,$. Equation (\ref{e14}) is analogous to
(\ref{e8}) for the Hermitian quartic theory, and can be confirmed analytically \cite{r7}. 

To calculate $r$ analytically we follow the procedure used above for the
Hermitian quartic theory. Define $g_p\equiv-i^n G_p/(p-1)!$ and rewrite the DS
equations for the Green's functions $G_{n}$ as one compact equation:
$$g_p=\textstyle{\frac{1}{p-1}\sum_{k=1}^{p-1}g_k g_{p-k}+
\frac{1}{2}\delta_{p,3}}\quad(p\geq2).$$
Next, multiply by $(p-1)x^p$, sum from $p=2$ to $\infty$, and define the
generating function $f(x)\equiv\sum_{p=1}^\infty x^p g_p$, which obeys the
Riccati equation $xf'(x)-f(x)=f^2(x)+x^3$.

Substituting $f(x)=-xu'(x)/u(x)$ linearizes this equation: $u''(x)=-xu(x)$.
This is an Airy equation whose general solution is $u(x)=a\,{\rm Ai}(-x)+b\,{\rm Bi}
(-x)$. From $f'(0)=g_1=-3^{1/3}\Gamma\big(\tthird\big)/\Gamma\big(\third\big)$
we find that $a$ is arbitrary and $b=0$, so $f(x)=x\,{\rm Ai}'(-x)/{\rm Ai}(-x)$.

The power series for the generating function $f(x)$ blows up when the
denominator vanishes, when $x=2.338\,107\,410\,460\,...$. This
is the radius of convergence of the series and its {\it inverse} 
is the value of $r$ in (\ref{e14}).

The rapid growth of $G_n$ in (\ref{e14}) explains the slow convergence and
inaccurate numerical results obtained by truncating the DS equations
(Fig.~\ref{f4}). Once again, using this asymptotic approximation
instead of setting $G_n=0$ gives extremely accurate and rapidly convergent
approximations to $G_1$ \cite{r7}.

\vspace{0.1cm}
\noindent{\bf Non-Hermitian quartic $D=0$ theory:} The Lagrangian $\cL=-\fourth
\phi^4$ defines a non-Hermitian $\cPT$-symmetric theory where
$$G_1=\textstyle{\frac{\int dx\,x\exp\big(x^4/4\big)}{\int dx\,\exp
\big(x^4/4\big)}=-\frac{2i\sqrt{\pi}}{\Gamma(1/4)}}
=-0.977\,741...\,i,$$
and the path of integration lies inside a $\cPT$-symmetric pair of Stokes
sectors in the lower-half complex-$x$ plane.

The first three DS equations are
\begin{eqnarray}
&&G_3=-G_1^3 -3 G_1 G_2,\nonumber\\
&&G_4=-3 G_1 G_3 -3 G_2^2 -3 G_1^2 G_2 -1,\nonumber\\
&&G_5=-3 G_1 G_4 -9 G_2 G_3 -3 G_1^2 G_3 -6 G_1 G_2^2.
\label{e15}
\end{eqnarray}
Solving these equations is harder than for the Hermitian quartic or the
non-Hermitian cubic theory, as {\it two} Green's functions must be set to zero
to close the system, and {\it two} coupled polynomials equations must be solved
simultaneously. The leading-order  truncation
leads to: $G_1=-i\,(3/2)^{1/4}=-1.106682...\,i$, which
differs from the exact $G_1$ above by $13.2\%$.

\begin{figure}[t]
\centering
\includegraphics[scale = 0.30]{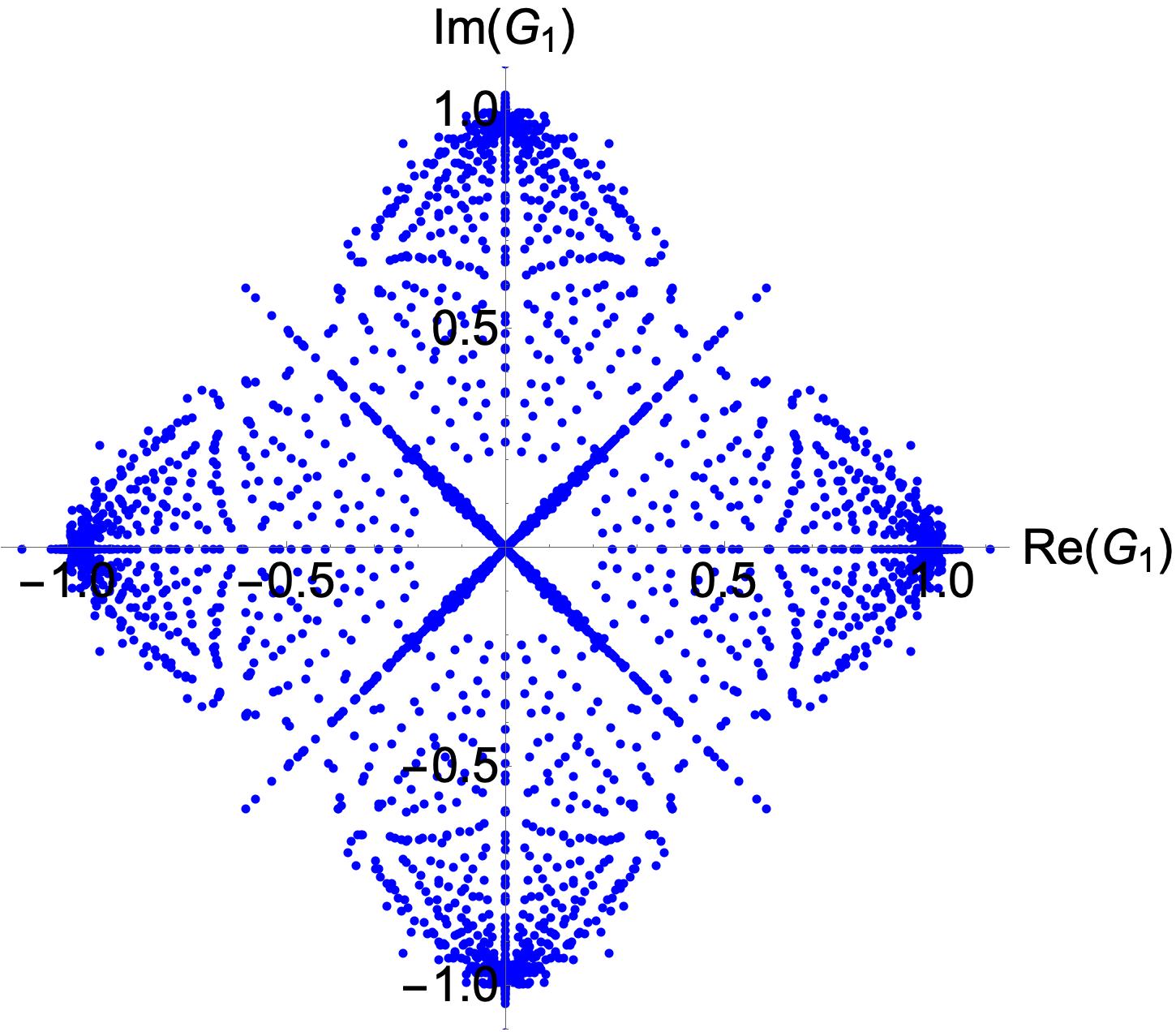}
\caption{All roots $G_1$ up to $n=33$ plotted as points in the complex plane.
The roots exhibit fourfold symmetry but only those on the negative-imaginary
axis respect $\cPT$ symmetry.
}
\label{f5}
\end{figure}

This procedure is continued for larger $n$. The number of roots increases rapidly
and the roots have fourfold symmetry in the complex plane. All roots up to $n=33$
are shown in Fig.~\ref{f5}. There are four concentrations of roots on the axes but 
$\cPT$ symmetry requires that $G_1$ be negative imaginary. Unlike Fig.~\ref{f4} the
dots are scattered over the complex plane because truncating the DS equations gives two {\it
coupled polynomial equations}.

We can determine the asymptotic behavior of $G_n$ for large $n$ from the DS
equations in (\ref{e15}). We find $G_n\sim-i(n-1)!\,(-i)^n r^n$, where $r=0.34640...
\,$. This result is analogous to the asymptotic behavior in (\ref{e14}).

\begin{figure}[t]
\centering
\includegraphics[scale = 0.28]{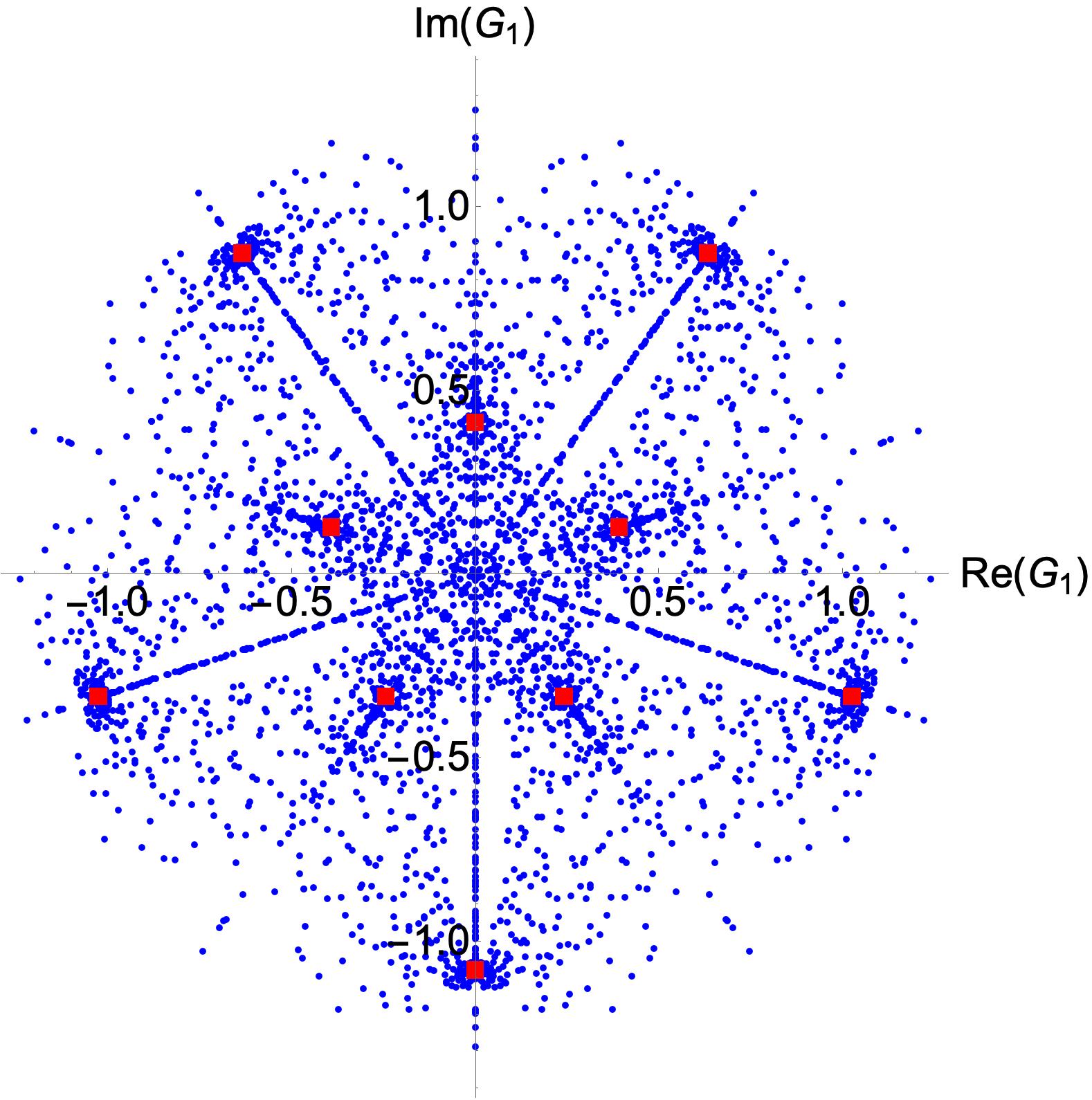}
\caption{Solutions to the DS equations for a quintic $D=0$ field theory. Exact values are 
denoted by squares.}
\label{f6}
\end{figure}

\vspace{0.1cm}
\noindent{\bf Quintic and sextic $D=0$ theories:} The DS equations for the
$\cPT$-symmetric $D=0$ Lagrangian $-\fifth i\phi^5$ require that {\it three}
higher Green's functions be set to 0 to close the truncated system, leading
to three coupled polynomial equations for $G_1$, $G_2$, and $G_3$. Going to
 the $n=11$ truncation we see {\it ten} concentrations of roots in Fig.~\ref{f6}. (The
DS equations are insensitive to the choice of Stokes sectors for the functional
integral.) There are {\it two} pairs of $\cPT$-symmetric boundary conditions,
which give rise to two imaginary values of $G_1= 0.412\,009...\,i $
and $G_1=-1.078\,653...\,i$\cite{r8}, seen on Fig.~\ref{f6} as heavy dots.

\begin{figure}[t]
\centering
\includegraphics[scale = 0.245]{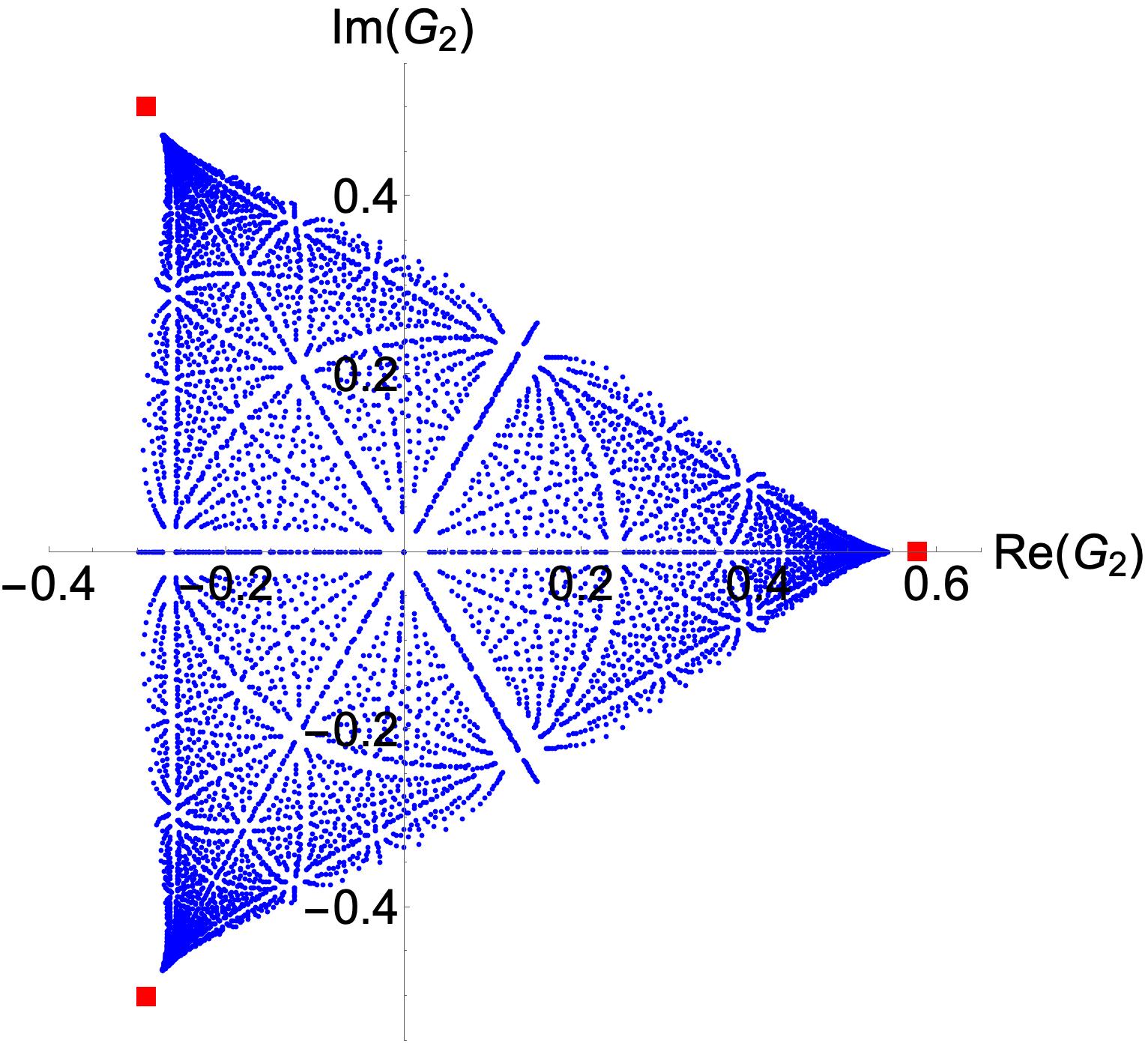}
\caption{Sextic case showing three concentrations of parity-symmetric solutions
for $G_2$ differing from the exact values (squares) by a few percent.}
\label{f7}
\end{figure}

For the sextic case $\cL=\sixth\phi^6$ we truncate the DS equations and set the
{\it four} highest Green's functions to 0. We must solve four coupled polynomial
equations. To reduce the number of solutions we impose parity symmetry, so $G_1
=G_3=0$. This eliminates all but three pairs of Stokes sectors. Figure~\ref{f7}
shows three concentrations of roots for $G_2$ up to the $n=32$ truncation. The
exact values of $G_2$ (squares) are $6^{1/3}\sqrt{\pi}/\Gamma(1/6)=0.578\,617
...$ and $-0.289\,302...\pm 0.501\,097\,i$; the error is a few percent.

\vspace{0.1cm}
\noindent{\bf Summary:} For five $D=0$ field theories we have shown that the
truncated DS equations yield underdetermined polynomial systems. There is no
effective strategy to solve such systems: Closing the systems by setting higher
Green's functions to zero gives sequences of approximants that converge to incorrect
limiting values. Replacing higher Green's functions with mean-field-like
approximations also gives incorrect limiting values, and this approach has the
drawback that if $D>0$, renormalization is required. The one numerically
accurate approach is to replace the higher $G_n$'s by their large-$n$ asymptotic
behaviors. This is difficult when $D>0$, but we believe that it may be possible to calculate, and it
presents an interesting avenue for further research.

This study emphasizes that the DS equations are {\it local}. Deriving the DS 
equations assumes only that the functional integrals {\it exist}; the DS equations 
are insensitive to which Stokes sectors in function space are used. As a result, 
the approximants try (but fail) to approach many different limits, most of which are
complex \cite{r9}.

The accuracy of the DS truncations worsens when interaction terms have higher
powers of the field because the indeterminacy of the system increases. More
Green's functions must be set to 0 to close the truncated system.

For Lagrangians having a weak-coupling constant $g$ we can expand all
$G_n$ in the DS equations as series in powers of $g$. This removes all
ambiguities discussed in here and gives the unique weak-coupling expansion
for each $G_n$. However, this merely replicates a Feynman-diagram calculation
of the Green's functions and totally ignores the nonperturbative
content of the theory. 

\acknowledgments
CMB thanks the Alexander von Humboldt and the Simons Foundations, and the UK
Engineering and Physical Sciences Research Council for financial support.



\clearpage

\vskip 1.0cm
\begin{widetext}

\begin{center}
\large{\textbf{Supplemental Material to "Underdetermined Dyson-Schwinger equations"}}
\end{center}

\section{Derivation of the DS equations for $D=1$ }

The DS equations follow
directly on differentiating the functional integral for $Z[J]$ (or $\log(Z[J])$)
with respect to $J$, giving $\gamma_n$ (or $G_n$), 
$$Z[J]=\textstyle{\int\cD\phi\,\exp\int dx}\{-\cL[\phi(x)]+J(x)\phi(x)\}.$$
($\cL$ is the Lagrangian, $J$ is a  c-number source, and $Z[0]$ is the 
Euclidean partition function.) 

For example, for a $D=1$ Hermitian quartic theory,
\begin{equation}
Z[J]={\textstyle\int} D\phi\,\exp\big[\textstyle{\int}dt(
-\half{\dot\phi}^2-\fourth\phi^4+J\phi)\big].
\label{e1}
\end{equation}
We take the VEV of the field equation $-{\ddot\phi}(t)+\phi^3(t)-J(t)=0$ and
divide by $Z[J]$:
\begin{equation}
-{\ddot G}_1(t)+\gamma_3(t,t,t)/Z[J]=J(t).
\label{e2}
\end{equation}
Note that $G_1(t)$ and $\gamma_3(t,t,t)$ are functionals of $J$.

We calculate $\gamma_3$ by differentiating $\gamma_1(t)=\langle 0|\phi(t)|0
\rangle=Z[J]G_1(t)$ twice with respect to $J(t)$:
\begin{eqnarray}
\gamma_2(t,t)&=&\langle 0|\phi^2(t)|0\rangle=Z[J]G_2(t,t)+Z[J]G_1^2(t),
\nonumber\\
\gamma_3(t,t,t)&=&\langle 0|\phi^3(t)|0\rangle=Z[J]G_3(t,t,t)\nonumber\\
&&\!\!\!+3Z[J]G_1(t)G_2(t,t)+Z[J]G_1^3(t).\nonumber
\end{eqnarray}
We then divide by $Z[J]$ and eliminate $\gamma_3$ in (\ref{e2}):
\begin{equation}
-{\ddot G}_1(t)+G_3(t,t,t)+3G_1(t)G_2(t,t)+G_1^3(t)=J(t).
\label{e3}
\end{equation}

We obtain the DS equations from (\ref{e3}) by repeated functional
differentiation with respect to $J$. For the first DS equation we set $J\equiv
0$. Parity invariance implies that odd-numbered Green's functions vanish, so the
first DS equation is trivial: $0=0$. For the second DS equation we differentiate
(\ref{e3}) with respect to $J(s)$ and set $J\equiv0$:
\begin{equation}
-{\ddot G}_2(s-t)+M^2 G_2(s-t)+G_4(s,t,t,t)=\delta(s-t),
\label{e4}
\end{equation}
where the renormalized mass is $M^2=3G_2(0)$.

Equation (\ref{e4}) is one equation in two unknowns, $G_2$ and $G_4$. (Each new
DS equation introduces one new unknown.) To proceed, we simply set $G_4=0$
in (\ref{e4}) and Fourier transform to get $(p^2+M^2){\tilde G}_2(p)=1$. The
inverse transform gives $G_2(t)=e^{-M|t|}/(2M)$, so $G_2(0)=1/(2M)$, yielding the 
solution is $M=(3/2)^{1/3}=1.145...\,$. This renormalized mass is the first excitation 
above the ground state and for this model $M=E_1-E_0=1.088...\,$. Thus, the 
leading-order DS result is 5.2\% high.

Next, we examine a $\cPT$-symmetric quartic theory in $D=1$; we change the sign
of the $\phi^4$ term in (\ref{e1}) \cite{r5}. The Green's functions are not
parity symmetric, so the odd-$n$ $G_n$'s do not vanish. The first DS
equation is nontrivial: $3G_1G_2(0)+G_1^3=0$. The second DS equation leads to
two equations: $M^2=-3[G_1^2+G_2(0)]$ and $G_2(0)=1/(2M)$. We solve these three
equations: $M=3^{1/3}=1.442...\,$. The exact value of $M$ obtained by solving the 
Schr\"odinger equation for the $\cPT$-symmetric quantum-mechanical Hamiltonian 
$H=\half p^2-\fourth x^4$ is $E_1-E_0=1.796...$, so the DS result is 19.7\% low.

These two examples motivate us to ask if higher truncations improve the
accuracy, but this leads to nonlinear integral equations requiring detailed
numerical analysis. However, we can solve the DS equations in high order when
$D=0$, which we do for five $D=0$ field theories.

\section{Hermitian quartic theory in D=0}

As observed in the main text, Eq.~(8), the asymptotic behavior of the Green's 
functions for the Hermitian quartic theory was determined to be
\begin{equation}
G_{2n} \sim 2r^{2n}(-1)^{n+1}(2n-1)!~~(n\to\infty),
\label{e8}
\end{equation}
where $r=0.409\,505\,7...\,.$. Clearly $G_{2n}$ grows rapidly with $n$, so that 
it is surprising that the procedure of truncation still leads to a relatively 
accurate result, as was shown in Fig.~1 of the main text. We have argued that 
this is because, while the exact value of the left hand side of the appropriate 
equation in Eq.~(6) is big, the terms on the right hand side are comparably big. 

The numerical technique of Legendre interpolation provides a useful
analogy \cite{r1}. Given $n$ data points $x_1,
\,...,\,x_n$ at which we measure a function $f(x)$, $f(x_1)=f_1,~...,~f(x_n)=
f_n$, one can fit this data with a polynomial $P_{n-1}(x)$ of degree $n-1$ that
passes exactly through the value $f(x_k)$ at $x=x_k$ for $1\leq k\leq n$. There
is a simple formula for this polynomial. The problem with this construction is
that between data points the polynomial exhibits huge oscillations where it
becomes alternately large and positive and large and negative due to a
fundamental instability associated with high-degree polynomials. To avoid this
instability, which is associated with the stiffness of polynomials, one can use
a {\it least-squares} fit, which passes near but not exactly through the input
data points. This is why {\it cubic} splines are used to approximate functions
rather than, say, octic splines. The instability associated with high-degree
polynomials allows the DS approach to work fairly well. If we use the exact
values of the Green's functions on the right side, we obtain the exact value of
the Green's function on the left side, which is a huge number. Evidently,
changing the Green's functions on the right side of Eq.~(6) of the main text
slightly by replacing the exact values by approximate values of the lower 
Green's functions now gives 0, instead of $G_{2n}$.

To determine the asymptotic behavior of $G_{2n}$ {\it analytically}, we defined a 
generating function, which, after manipulation, leads to the expression for 
$y(x)$ given in Eq.~(10) of the main text, 
$$
y(x)=\textstyle{\frac{2\sqrt{2}}{\Gamma(1/4)}\int_0^\infty}dt\,\cos(xt)\,
e^{-t^4/4}.
$$
We plot $y(x)$ in Fig.~\ref{f1}, 
from which one can determine the zero nearest to the origin. This lies at 
$x_0=\pm2.4419682...$, and has the inverse 0.409506\dots, corresponding 
to the value of $r$ found numerically through Richardson extrapolation.

\begin{figure}[h]
\centering
\includegraphics[scale = 0.25]{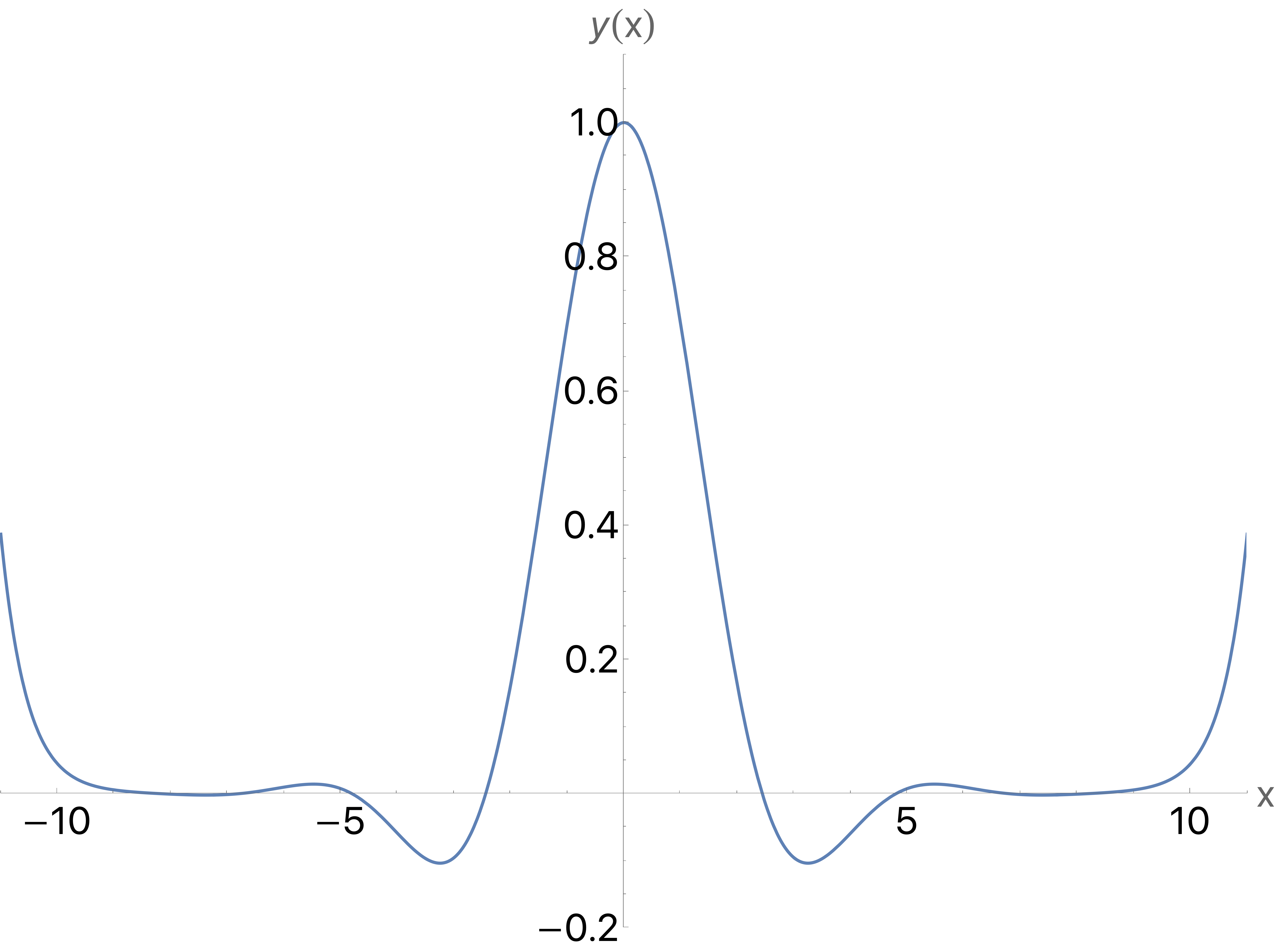}
\caption{Plot of $y(x)$ normalized to $y(0)=1$. The function $y(x)$ is even. 
The zero nearest to the origin lies at $x_0=\pm2.4419682...$ the inverse of this 
number is 0.409506\dots, which confirms the numerical result for $r$ below 
Eq.~(8) of the main text.}
\label{f1}
\end{figure}

\section{non-Hermitian cubic theory in $D=0$}

We first comment that the roots in Fig.~2 of the main paper have threefold symmetry 
because the truncated DS equations give polynomials having only powers of $x^3$ 
(apart from a root at 0). The DS equations depend only {\it locally} on the integrand of the functional
integral; they are totally insensitive to the boundary conditions on the
functional integrals. There are three pairs of Stokes sectors of angular opening
$60^\circ$ inside of which the integration path can terminate.
These sectors are centered about $\theta_1=\frac{\pi}{2}$, $\theta_2=-\frac{\pi}
{6}$, or $\theta_3=-\frac{5\pi}{6}$. If the integration path terminates in the
\cPT-symmetric (2,3) sectors, $G_1$ is negative imaginary, but if it terminates
in the (1,2) or (1,3) sectors, $G_1$ is complex.

Secondly, using the asymptotic approximation to $G_n$ given in Eq.~(13) of the main text
to calculate the successive orders of approximation to $G_1$ leads to extremely
accurate and rapidly convergent values of $G_1$. As we are only interested in 
solutions along the negative imaginary axis, in Fig.~\ref{f2} we show a sector of the
complex plane calculated in this way, together with the standard truncation 
scheme $G_n=0$. This can be compared with the inset in Fig.~3 of the main 
text. Using the asymptotic expansion evidently leads to many solutions in the 
complex plane. In addition, an accurate value of $G_1$ appears on the 
imaginary axis.  This can be best seen by plotting the absolute values of 
$G_1$ as a function of $n$, and which is presented in Fig.~3. In addition to the 
solution lying numerically above the absolute value of $G_1$, the exact solution
also emerges.

\begin{figure}[h]
\centering
\includegraphics[scale = 0.32]{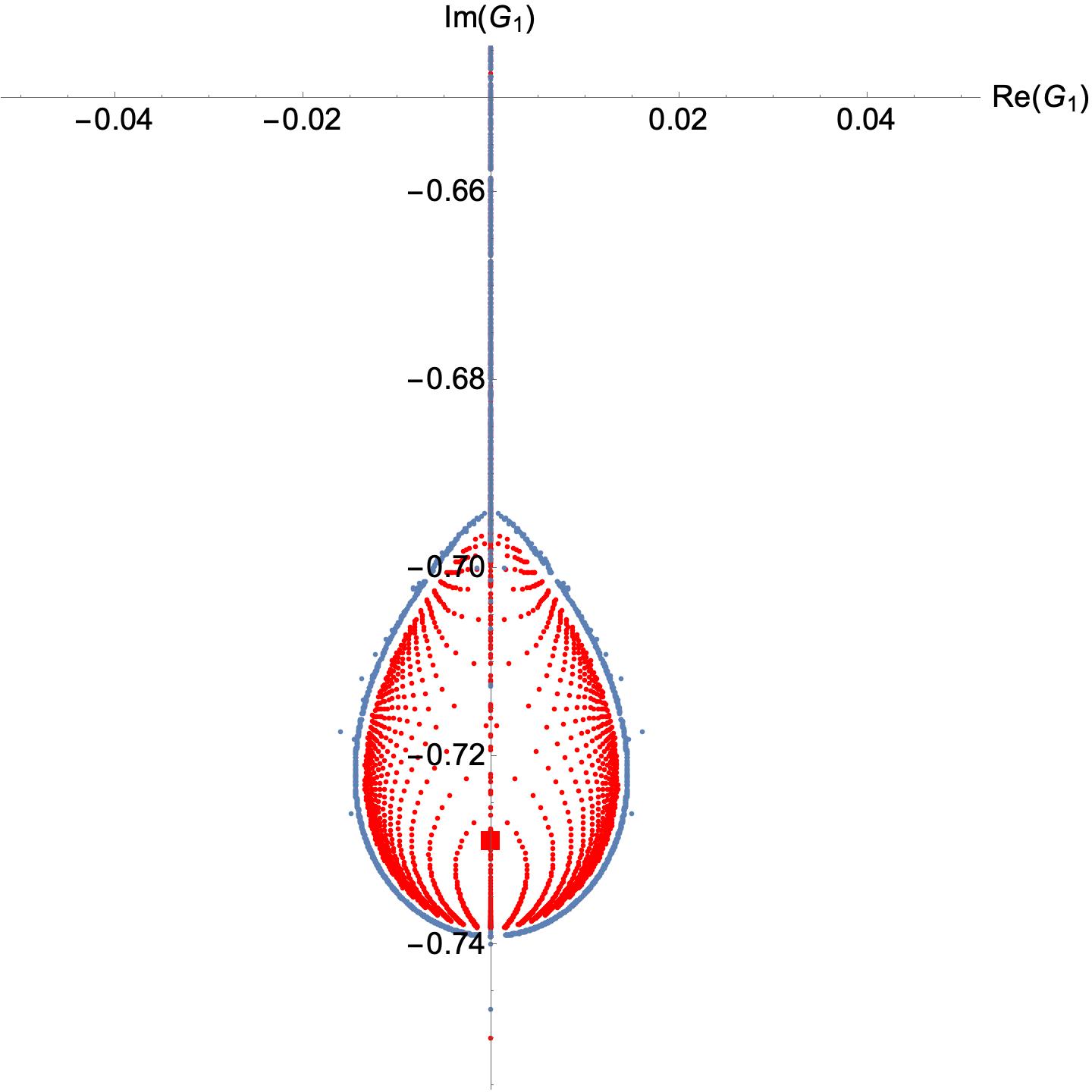}
\caption{Solutions of the truncated DS equations (12) of the main text, for the
non-Hermitian cubic theory, corresponding to Fig.~3 of the main text. In addition to the
solution shown there, we include solutions based on the asymptotic behavior of $G_n$
(internal structure in the loop). This leads to a distribution pattern in the complex plane, 
as well as a rapid convergence to the exact value of $G_1 = -0.72901113\dots i$, which is
indicated by the square.
}
\label{f2}
\end{figure}

\begin{figure}[h]
\centering
\includegraphics[scale = 0.45]{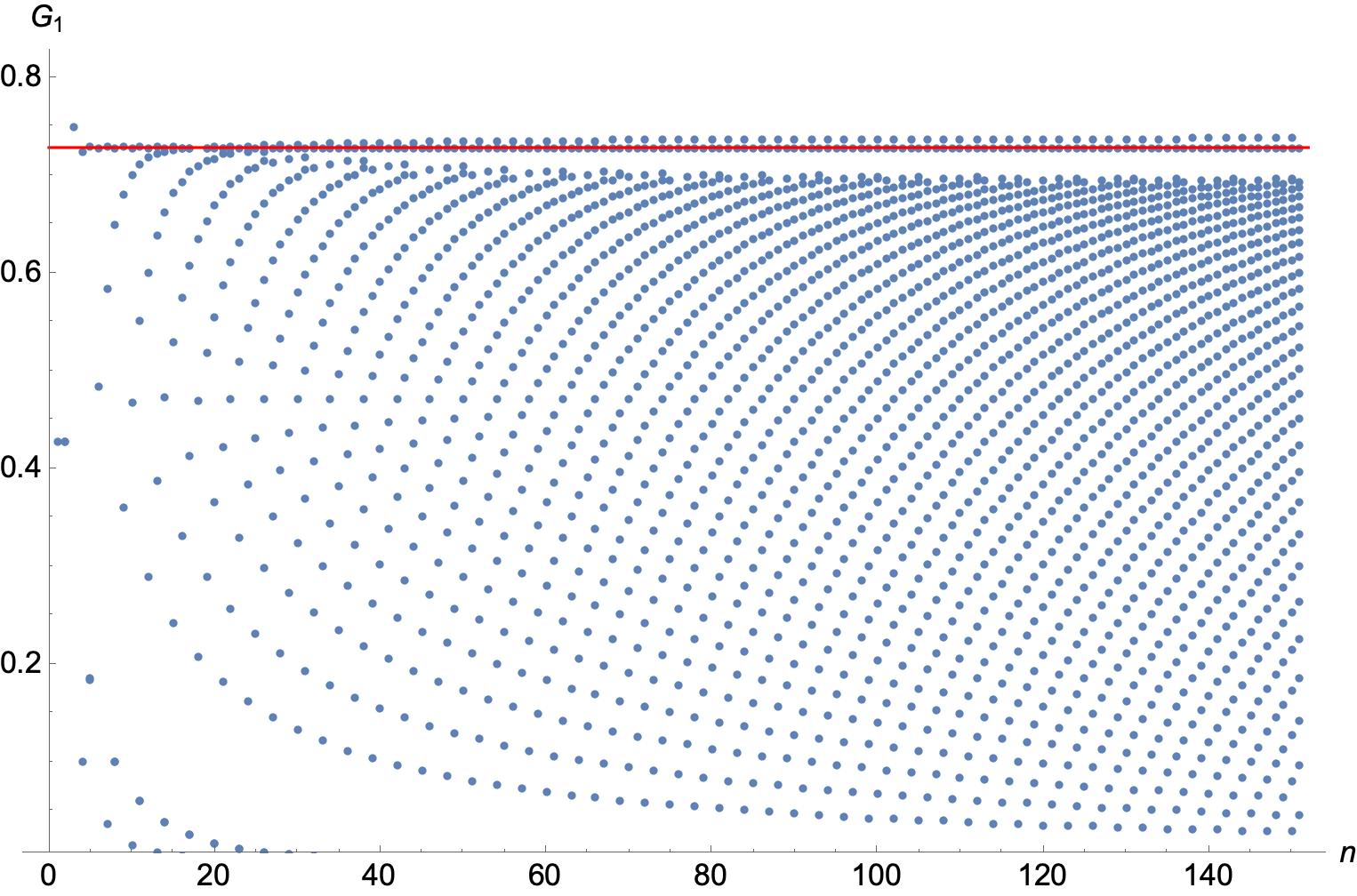}
\caption{Absolute values of the solution to the truncated DS equations (12) 
assuming an asymptotic form for $G_n$, as given in Eq.~(13) of the main 
text for $n$ up to 150. Rapid convergence to the exact value of 
$|G_1| = 0.72901113\dots$, indicated by the heavy line, is observed. 
}
\label{f3}
\end{figure}

\end{widetext}



\begin{thebibliography}{99}
\bibitem{r1} We calculate $G_n$, rather than $\gamma_n$, to avoid vacuum
divergences. Disconnected contributions to $\gamma_n$ introduce factors of the
spacetime volume, which is infinite.
\bibitem{r9} The quantum-mechanics
analog of this problem is that the spectrum of a Hamiltonian is undetermined
until the boundary conditions on the Schr\"odinger eigenfunctions are given.
For example, the spectrum of the harmonic oscillator Hamiltonian is
either positive or negative depending on whether the eigenfunctions vanish as
$x\to\pm\infty$ or $x\to\pm i\infty$. See Ref. 5.
\end{thebibliography}
\end{document}